%% file: loop.tex
\documentclass[pre,twocolumn,groupedaddress,showpacs,showkeys]{revtex4}

			\usepackage{amssymb}
			\usepackage{amsmath}
			\usepackage{graphicx}
			\usepackage{subfigure}
			\usepackage{dsfont}
			\usepackage{mathrsfs}
			\usepackage{natbib}       
			\usepackage[dvips]{color} 
			\input loopDefs	  
\begin{document}
				\title{
Variational method for finding periodic orbits in a general flow
				}\author{
Yueheng Lan			}\email{gte158y@prism.gatech.edu}\author{
Predrag Cvitanovi\'{c}		}\email{predrag.cvitanovic@physics.gatech.edu}
\affiliation{
		Center for Nonlinear Science, School of Physics,\\
		Georgia Institute of Technology, Atlanta 30332-0430, U.S.A}
				\date{\today}

				\begin{abstract}
A variational principle for determining unstable periodic orbits of 
flows as well as unstable spatio-temporally periodic solutions of 
extended systems is proposed and implemented. An initial loop approximating
a periodic solution is evolved in the space of loops toward a true periodic solution
by a minimization of local errors along the loop. The 
``\descent'' partial differential 
equation that governs this evolution is an infinitesimal step version of
the damped Newton-Raphson iteration. The feasibility of the method is demonstrated by its application to
the H\'enon-Heiles system, the circular restricted three-body problem,
and the Kuramoto-Sivashinsky system in a  weakly turbulent regime.		
\end{abstract}
					\maketitle
\section{Introduction}

The periodic orbit theory of classical and quantum chaos\rf{ruelle,gutbook} 
is one of the major advances in the study of long-time behavior of chaotic 
dynamical systems. 
The theory expresses all long time averages over chaotic
dynamics in terms of cycle expansions\rf{cexp},
sums over
periodic orbits  (cycles)  ordered hierarchically according to the orbit length, stability,
or action. 
If the symbolic dynamics is known, and the flow is
hyperbolic, longer cycles are shadowed by the shorter ones, and
cycle expansions converge  exponentially or even 
super-exponentially with the cycle length\rf{hhrugh92}.

A variety of 
methods for determining all periodic orbits up to a given length 
have been devised and successfully implemented 
for low-dimensional systems\rf{DasBuch,decar2,lfind,afind,mfind}.
For more complex dynamics, such as turbulent flows\rf{frisch}, nonlinear
waves\rf{cgl}, or quantum fields\rf{gauge,chfield} with high 
(or infinite) dimensional phase spaces and complicated dynamical behavior,
many of the existing methods become unfeasible in practice. 
In the most computationally demanding calculation carried out so far, Kawahara
and Kida\rf{KawKida01} have found two periodic solutions in a $15,422$-dimensional
discretization of a turbulent plane Couette flow.
The topology of high-dimensional flows is hard to visualize, and 
even with a decent starting guess for the shape and location of a periodic
orbit, methods like the Newton-Raphson method are likely to fail. In 
\refref{CvitLanCrete02} we have argued that variational, cost-function minimization 
methods offer a robust alternative. 
Here we derive, implement and discuss in detail one
such new variational method for finding periodic orbits in
general flows, and specifically high-dimensional flows.

In essence, any numerical algorithm for finding periodic orbits is based on
devising a new dynamical system which possesses the desired orbit as an
attracting fixed point with a sizable basin of attraction. Beyond that, there is 
much freedom in constructing such system. 

For example, the multipoint shooting method eliminates the
long-time exponential
instability of unstable orbits by splitting an orbit into a number of short segments, each with a
controllable expansion rate. 
The multiple shooting 
combined with the Newton-Raphson method is an efficient tool for
locating periodic orbits of maps\rf{ChristiansenDasBuch}. 
A search for  periodic orbits of a continuous time flow can be reduced to
a multiple shooting search for  periodic orbits of a set of maps
by constructing
a set of phase space Poincar\'{e} sections  such that an orbit leaving
one section reaches the next one in
a qualitatively predictable manner, without traversing other sections
along the way. In turbulent, high-dimensional
flows such sequences of sections are hard to come by.
One solution might be a large set of Poincar\'{e} sections, 
with the intervening flight segments short and controllable. 

Here 
we follow a different  strategy, and discard Poincar\'{e} sections altogether;
we replace maps between spatially
fixed Poincar\'{e} sections, by maps induced by discretizing the time
evolution into small time steps. For sufficiently small  
time steps such maps are small deformations of identity. 
We distribute many points along a smooth loop $\Loop$, our 
initial guess of
a cycle location and its topological layout. If both
the time steps and the loop deformations are taken infinitesimal, 
a partial differential equation
 governs the  ``\descent'', a fictitious time  flow of a trial loop $\Loop$
into a genuine
cycle $p$, with exponential convergence in the fictitious time
variable. 
We then use methods developed for solving
PDEs to get the solution. Stated succinctly, the idea of our method
is to make an informed rough guess of what the desired cycle
looks like globally, and then use a variational method to drive the
initial guess towards the exact solution. For robustness, we replace the
guess of a single orbit point by a guess of an entire orbit. For
numerical safety we replace the Newton-Raphson iteration  by the 
``\descent'', a differential flow that minimizes a {\costFct} computed as
deviation of the approximate flow from the true flow along a smooth loop
approximation to a cycle. 
    
    In \refsect{sec:der} we derive the partial differential equation
which governs the evolution of an initial guess loop toward a cycle and
the corresponding {\costFct}. An extension of the method to Hamiltonian
systems and systems with higher time derivatives is presented in \refsect{sec:ext}.
Simplifications due to symmetries and details of our
numerical implementation of the method are discussed in \refsect{sec:nm}. 
In \refsect{sec:applt} we test
the method on the H\'{e}non-Heiles system, the
restricted three body problem, and  a weakly turbulent 
Kuramoto-Sivashinsky system. 
We summarize our results and 
discuss possible improvements of the method in \refsect{sec:sum}.

\section{The \descent\  method in loop space}
\label{sec:der}

\subsection{A variational equation for the loop evolution}

A periodic orbit is a
solution $(\pSpace,\period{})$, $\pSpace \in \reals^{d}$,
$\period{} \in \reals$ of the {\em periodic orbit condition}
\beq
f^{\period{}}(\pSpace) = \pSpace
\,,\qquad \period{} > 0
\label{e:periodic}
\eeq
for a given flow or discrete time mapping $x \mapsto f^t({x})$.
Our goal is to determine periodic 
orbits of flows defined by first order ODEs
\beq
\frac{d\pSpace}{dt}=v(\pSpace)
	\,,\qquad
 \pSpace \in \pS \subset \mathbb{R}^d
	\,,\qquad 
 (\pSpace,v) \in \bf{T}\pS  
\label{fl}
\eeq
in many (even infinitely many) dimensions $d$. Here $\pS$ is the phase space 
(or state space) in which evolution takes place, 
$\bf{T}\pS$ is the tangent bundle\rf{arnold92},
and the vector field $v(\pSpace)$ is assumed to be smooth (sufficiently 
differentiable) almost everywhere.

We make our initial guess at the shape  and the location of a cycle $p$ by
drawing a loop $\Loop$, a smooth, differentiable 
closed curve $\lSpace(s)\in \Loop \subset \pS$, where 
$s$ is a loop parameter. As the loop is periodic, we find it convenient
to restrict $s$ to  $[0,2\pi]$, with the periodic condition 
$\lSpace(s)=\lSpace(s+2\pi)$. 
Assume that $\Loop$ is close to the true cycle $p$,
pick $N$ pairs of nearby points 
along the loop and along the cycle
\bea
\lSpace_n &=& \lSpace(s_n)
	\,,\qquad
0 \leq s_1 < \ldots < s_N < 2\pi 
        \,,	\continue
\pSpace_n &=& \pSpace(t_n)
	\,,\qquad
 0 \leq t_1 < \ldots <  t_N < \period{p} 
\,,
\label{LoopDiscret}
\eea
and denote by $\delta \lSpace_n$ the deviation of a point $\pSpace_n$ on the 
periodic orbit $p$ from the nearby point $\lSpace_n$, 
\[
\pSpace_n  = \lSpace_n+\delta \lSpace_n
\,.
\] 
The deviations $\delta \lSpace$ are assumed small, vanishing as
$\Loop$ approaches $p$.

The orientation of the $s$-velocity vector tangent to the loop $\Loop$
\[
\lVeloc(\lSpace)=\frac{d \lSpace}{ds}\, 
\]
is intrinsic to the loop, but its
magnitude depends on the (still to be specified)
parametrization $s$ of the loop.

At each loop point $\lSpace_n \in \Loop$ we thus have two vectors,
the loop tangent
$\lVeloc_n=\lVeloc(\lSpace_n)$ and the flow velocity 
$v_n=v(\lSpace_n)$. Our goal is to deform $\Loop$ until
the directions of $\lVeloc_n$ and $v_n$ coincide for all $n=1,\ldots,N$, 
$\; N\to \infty$, {\ie} $\Loop = p$. To match their magnitude, we introduce a
local time scaling factor 
\begin{equation}
\lambda(s_n)  \equiv \Delta t_n/\Delta s_n\,, 
\label{dt}
\end{equation}
where $\Delta s_n=s_{n+1}-s_n, n=1,\ldots,N-1 \,,\Delta s_N=2\pi-(s_N-s_1)$,
and likewise for $\Delta t_n$. The scaling factor
$\lambda(s_n)$ 
ensures that the loop increment $\Delta s_n$ is 
proportional to its counterpart $\Delta t_n + \delta t_n$ 
on the cycle when the loop $\Loop$ is close to the
cycle $p$, with $\delta t_n \to 0$ as $\Loop \to p$. 

Let 
$ \pSpace(t) = \flow{t}{\pSpace} $
be the state of the system at time $t$ obtained by
integrating \refeq{fl}, and 
$\jMps(\pSpace,t) = d \pSpace(t)/d \pSpace(0)$ be
the corresponding {\jacobianM} obtained by integrating
\beq
\frac{d\jMps}{dt} = \Mvar \jMps
\,,\quad
\Mvar_{ij}=\frac{\partial v_i}{\partial x_j}
\,,\qquad
\mbox{with } \jMps(x,0)=\mathds{1} 
\,. 
\label{doa}
\eeq
 Since the point
$\pSpace_n=\lSpace_n+\delta \lSpace_n$ is on the cycle,
 \begin{equation}
\flow{\Delta t_n+\delta t_n}{\lSpace_n+\delta \lSpace_n}=\lSpace_{n+1}+
\delta \lSpace_{n+1} \label{bd}
\,.
\end{equation}
 Linearization 
\[
\flow{\delta t}{\pSpace} \approx \pSpace + v(\pSpace) \delta t
        \,, \qquad
\flow{t}{\pSpace+\delta\pSpace} \approx \pSpace(t) + \jMps(\pSpace,t) \delta\pSpace
        \,, 
\]
of  \refeq{bd} about the loop point $\lSpace_n$ and the time interval 
$\Delta t_n$ to the next cycle point
leads to the multiple shooting 
Newton-Raphson equation, for any step size  $\Delta t_n$:
\begin{equation}
\delta \lSpace_{n+1}-\jMps(\lSpace_n,\Delta t_n) \delta \lSpace_n -v_{n+1}
\delta t_n=\flow{\Delta t_n}{\lSpace_n}-\lSpace_{n+1} 
\,.
\label{bd1}
\end{equation}

Provided that the initial guess is sufficiently good, the Newton-Raphson iteration 
of \refeq{bd1} generates a sequence of loops $\Loop$
with a  decreasing {\costFct}\rf{CvitLanCrete02}
 \begin{equation}
 \costF(\lSpace) \equiv \frac{N}{(2\pi)^2}
                          \sum_{i=1}^{N}(\flow{\Delta t_n}{\lSpace_n}-\lSpace_{n+1})^2, 
\qquad \lSpace_{N+1}=\lSpace_1
\,. 
\label{cost}
\end{equation}
The prefactor ${N/(2\pi)^2}$ makes the definition of $\costF$ consistent with
\refeq{funal} in the $N \to \infty$ limit.
If the flow is locally strongly unstable,  
the neighborhood  in which the linearization is  valid
could be so small that the full Newton step
would overshoot, rendering ${\costF}$ bigger rather than smaller. In this
case the step-reduced, damped Newton method is needed. As proved 
in~\refref{kellbv}, under conditions 
satisfied here, $\costF$ decreases monotonically if
appropriate step size is taken.
If infinitesimal steps are taken, decrease of ${\costF}$ is ensured.
We parametrize such continuous deformations of the loop by a 
{\em fictitious time} $\tau$. 

\begin{figure}[t] 
\centering
(a)~~\includegraphics[width=4.0cm]{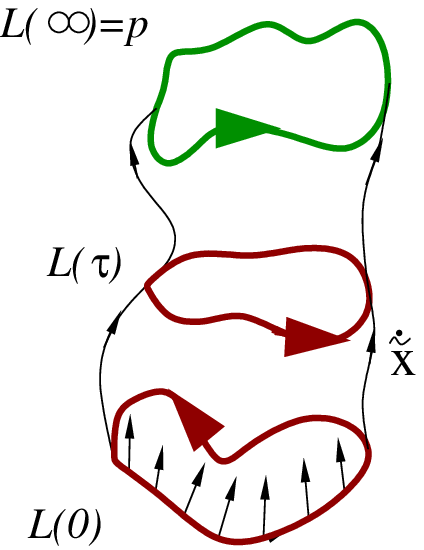}
\hspace{0.2in}
(b)~\includegraphics[width=5cm]{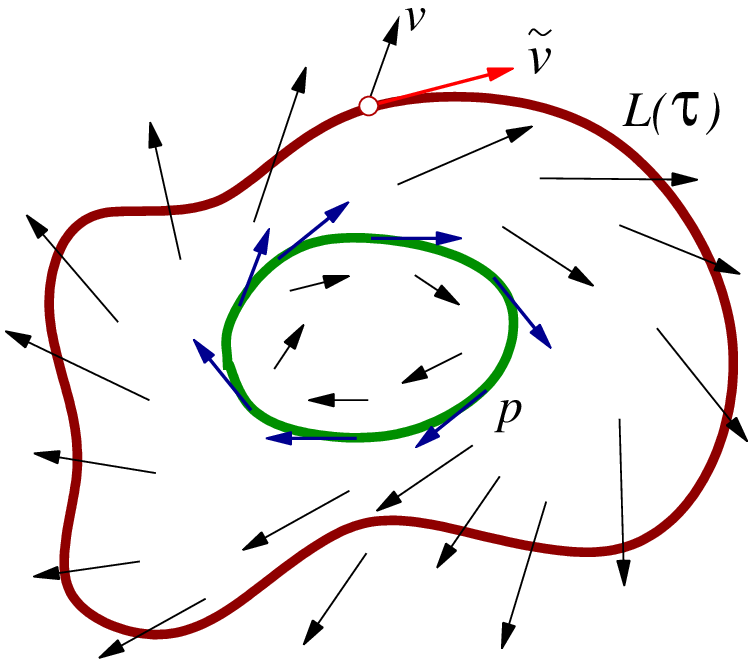}
\caption{
(a) An annulus $\Loop(\tau)$ swept by the {\descent} flow $d{\lSpace}/d\tau$,
 connecting 
smoothly the 
initial loop $\Loop(0)$ to the periodic orbit $p=\Loop(\infty)$. (b) In general the loop
velocity field $\lVeloc(\lSpace)$ does not coincide with $\lambda v(\lSpace)$; for a periodic 
orbit $p$, it does so at every $x \in p$.        }
\label{f:velocField}
\end{figure}

We fix $\Delta s_n$ and proceed by $\delta \tau$ each step of the iteration,
{\ie}, multiply the right hand side of 
\refeq{bd1} by $\delta \tau$.
According to \refeq{dt}, the change
of $\Delta t_n$ with respect to $\tau$ is equal to $\delta
t_n=\frac{\partial \lambda}{\partial \tau}(s_n,\tau) \delta \tau \Delta
s_n$. As $\delta \lSpace_n=\frac{\partial}{\partial \tau} 
\lSpace(s_n,\tau) \delta \tau$, 
dividing both sides of \refeq{bd1} by $\delta \tau$ yields
 \bea
	&&
\frac{d \lSpace_{n+1}}{d \tau}-\jMps(\lSpace_n,\Delta t_n) \frac{d \lSpace_n}
{d \tau} -v_{n+1}\frac{\partial \lambda}{\partial \tau}(s_n,\tau) 
\Delta s_n
	\continue
	&& \qquad\qquad\qquad\qquad\qquad
     =\flow{\Delta t_n}{\lSpace_n}-\lSpace_{n+1} 
\,.\label{bd2}
\eea
 In the $N\rightarrow \infty$ limit, the stepsizes $\Delta s_n, \Delta
t_n= O(1/N)\rightarrow 0$, and we have
 \begin{eqnarray*}
v_{n+1}&\!\approx\!&v_n 
	\,, \qquad \qquad \qquad
\lSpace_{n+1} \approx \lSpace _n+\lVeloc_n \Delta s_n 
	\,,\\
\jMps(\lSpace_n,\Delta t_n)\!&\!\approx \!&\! 1+\Mvar(\lSpace_n) \Delta t_n
	\,,\quad
\flow{\Delta t_n}{\lSpace_n} \approx \lSpace_n+v_n \Delta t_n 
\,.
\end{eqnarray*}
 Substituting into 
\refeq{bd2} and using the scaling relation \refeq{dt},
  we obtain
 \begin{equation}
\frac{\partial^2 \lSpace}{\partial s \partial \tau}-\lambda \Mvar  
\frac{\partial \lSpace}{\partial \tau}-v\frac{\partial \lambda}{\partial \tau}
=\lambda v-\lVeloc. 
\label{bd3}
\end{equation}
 This PDE, which describes the evolution 
of a loop  $\Loop(\tau)$ 
toward a periodic
orbit $p$, is the central result of this paper. 
The family of loops so generated is parametrized
by $\lSpace=\lSpace(s,\tau)\in \Loop (\tau)$, where $s$ denotes the 
position along 
the loop,  and the fictitious time $\tau$ parametrizes the deformation of the loop, 
see \reffig{f:velocField}(a).
We refer to this infinitesimal step
version of the damped Newton-Raphson method as the ``\descent''.

The important feature of this equation is that a decreasing cost functional exists.
Rewriting \refeq{bd3} as
 \begin{equation}
\frac{\partial}{\partial \tau}(\lVeloc-\lambda v)
=-(\lVeloc-\lambda v) \,,
\label{cc0}
\end{equation}
we have 
\begin{equation}
\lVeloc-\lambda v= 
	e^{-\tau}(\lVeloc-\lambda v)|_{\tau=0} 
\,,
\label{cc}
\end{equation}
so the fictitious time $\tau$ flow decreases the {\costFct}al
 \beq
\costF[\lSpace]=\frac{1}{2 \pi}\oint_{\Loop(\tau)} ds \,
          \left(\lVeloc(\lSpace)-\lambda v(\lSpace)\right)^2 
\label{funal}
\eeq
monotonically as the loop evolves toward the cycle.

At each iteration step the differences of the loop tangent velocities and  
the dynamical flow velocities are reduced by the {\descent}. 
As $\tau \rightarrow \infty$, the fictitious time flow alignes
the loop tangent $\lVeloc$  with 
the dynamical flow vector $\lVeloc=\lambda v$, and the loop 
$\lSpace(s,\tau) \in \Loop(\tau)$, see \reffig{f:velocField}(b),
converges to a genuine periodic orbit $p = \Loop(\infty)$ of
 the dynamical flow $\dot{\pSpace}=v(\pSpace)$.  
Once the cycle $p$ is reached, by \refeq{dt}, 
$\lambda(s,\infty)=\frac{dt}{ds}(\lSpace(s,\infty))$, and the cycle period is given by 
 \[
\period{p}=\int_0^{2\pi} \lambda(\lSpace(s,\infty))ds
\,. 
\]
 Of course, as at this stage we have already identified the cycle, we may 
pick instead an initial point 
on $p$ and  calculate the period by a direct integration of the 
dynamical equations \refeq{fl}.

\subsection{Marginal directions and accumulation of loop points}
\label{sect:marg}

Numerically, two perils lurk in a direct implementation of the {\descent} 
\refeq{bd3}.
 
First, when a cycle is reached, it remains a cycle under a 
cyclic permutation of the representative points, so on the cycle
the operator 
\begin{eqnarray*}
\bar{A}=\frac{\partial}{\partial s}-\lambda A 
\end{eqnarray*}
has a marginal eigenvector $v(\lSpace(s))$ with 
eigenvalue $0$. 
If $\lambda$ is fixed, as the loop approaches the cycle, \refeq{bd3} 
approaches its limit
\[
\bar{A}\,\frac{\partial x}{\partial \tau}=0 \,. 
\]
Therefore, on the cycle, the operator $\bar{A}^{-1}$ becomes
singular and the numerical woes arise. 

The second potential peril hides in the freedom of choosing the loop
(re-)parametrization. Since $s$ is related to the time $t$
by the yet unspecified factor $\lambda(s,\tau)$, uneven distributions of the sampling
points over the loop $\Loop$ could arise, with the numerical discretization 
points $\lSpace_n$ 
clumping densely along some segments of $\Loop$ and leaving big gaps elsewhere, 
thus degrading the numerical smoothness of the loop. 

We remedy these difficulties by imposing constraints on 
\refeq{bd3}. In our calculation for Kuramoto-Sivashinsky system 
of \refsect{sec:applt}, the first 
difficulty is dealt with by introducing one Poincar\'{e} section, for example, 
by fixing one coordinate of one of the sampling points, 
$\lSpace_1(s_1,\tau)=const$.  
This breaks the translational invariance along the cycle. 
Other types of constraints might be better suited to a specific 
problem at hand. For example, we can demand that the average 
displacement of the sampling points along the loop vanishes, thus avoiding
a spiraling descent towards the desired cycle.

We deal with the second potential difficulty by choosing a particularly
simple loop parametrization.
So far, the
parametrization $s$ is arbitrary and there is much freedom in choosing the best
one for our purposes.
We pick $s-$ and $\tau-$independent constant scaling 
$\lambda(s, \tau) = \lambda$. 
With uniform grid size $\Delta s_n=\Delta s$ and fixed $\lambda$,
the loop parameter $s=t/\lambda$ is proportional to time $t$, and 
the discretization \refeq{bd3} distributes the sampling points
along the loop evenly in time.
As the loop approaches a cycle,
$\frac{\partial \lSpace}{\partial \tau}$ is numerically obtainable from 
\refeq{bd3}, and on the cycle the period is given by $\period{p}=2\pi \lambda$.

Even though this paper focuses on searches for periodic orbits, the {\descent} is
a general method. With appropriate modifications of boundary conditions and 
scaling of time,
\refeq{bd3} can be adapted to determination of homoclinic or heteroclinic 
orbits
between equilibrium points or periodic orbits of a flow, or more 
general boundary value
problems. Applied to 2-point boundary value problems, {\descent} is similar 
to the quasilinearization\rf{bl} but has
the advantage that the free parameter $\lambda(s,\tau)$ is available 
for adjusting scales in the problem and that searches
can be restricted to phase space submanifolds of interest. 
A simple example of a restriction to a submanifold are searches for cycles of 
a given energy, 
constrained to the $H(q,p)=E$ energy shell 
in the phase space of a Hamiltonian system. Furthermore, 
as we shall show now, the symplectic structure of Hamilton's equations 
greatly reduces the dimensionality of the submanifold that we need to consider.  

\section{Extensions of \descent}
\label{sec:ext}

    In classical mechanics
particle trajectories are also solutions of a variational principle, 
the Hamilton's variational principle.
For example, one can determine a periodic orbit of a 
billiard by wrapping around a rubber band of a roughly correct topology, and then
moving the points along the billiard walls until the length ({\ie}, the action) 
of the
rubber band is extremal (maximal or minimal under infinitesimal changes of the
boundary points). Note that the 
extremization of action requires only 
$D$ configuration coordinate variations,
not the full $2D$-dimensional phase space variations.

Can we exploit this property of the Newtonian mechanics to reduce the dimenionality
of  our  variational calculations?
The answer is yes, and easiest to understand in terms of the Hamilton's
variational principle which states that classical trajectories are extrema
of the Hamilton's principal function (or, for fixed energy $E$,  the action $S=R+Et$)
\[
R(q_1,t_1;q_0,t_0) = \int_{t_0}^{t_1} \! dt \,{\cal L}(q(t), {\dot q}(t),t)
\,,
\]
where
$ {\cal L}(q, {\dot q},t)$ is the Lagrangian.
Given a loop $\Loop(\tau)$ we can compute
not only the tangent ``velocity'' vector $\tilde{v}$, but also 
the local loop curvature or ``acceleration'' vector 
\[
\tilde{a} = 
\frac{\partial^2 \lSpace}{\partial s^2}
\,,
\]
and indeed, as many $s$ derivatives as needed. 
Matching the dynamical acceleration $a(\lSpace)$ (assumed to be functions of 
$\lSpace$ and $v(\lSpace)$) 
with the loop ``acceleration'' $\tilde{a}(\lSpace)$
results in a new {\costFct} and the corresponding PDE \refeq{cc0} for the evolution of the loop
\[
\frac{\partial}{\partial \tau}(\tilde{a}-\lambda^2 a)=-(\tilde{a}-\lambda^2 a)
\,.
\]   
We use $\lambda^2$ instead of $\lambda$ in order to keep the notation
consistent with \refeq{dt}, {\ie} $t=\lambda \,s$. 
  Expressed 
 in terms of the loop variables $\lSpace(s)$, the above equation becomes
 \bea
	&&
 \frac{\partial^3
\lSpace}{\partial^2 s \partial \tau}-\lambda \frac{\partial a}{\partial v} 
\frac{\partial^2 \lSpace}{\partial s \partial \tau}-\lambda^2 \frac{\partial a}
{\partial \lSpace} 
\frac{\partial \lSpace}{\partial \tau}
 +\left(\frac{\partial a}{\partial v} 
           \frac{\partial \lSpace}{\partial s}-2\lambda a
   \right)\!
             \frac{\partial \lambda}{\partial \tau}
	\continue
	&& \qquad\qquad\qquad\qquad\qquad
=\lambda^2 a-\tilde{a} \, ,
\label{hfl}
\eea 
where $v=\frac{\partial \lSpace}{\lambda \partial s}$. Although \refeq{hfl}
looks more complicated than \refeq{bd3}, in numerical fictitious time integrations, we
are rewarded by having to keep only half of the phase space variables.

    More generally, if a differential equation has the form:
\begin{equation}
\pSpaceDer{m}=f(\pSpace,\pSpaceDer{1},\cdots,\pSpaceDer{m-1}) \, , \label{gf}
\end{equation}
 where $\pSpaceDer{k}=\frac{d^k \pSpace}{dt^k},\, k=1,\cdots,m$ and $\pSpace \in 
\mathbb{R}^d$,
the same technique can be used to match the highest derivatives
$\lambda^m \pSpaceDer{m}$ and $\lSpaceDer{m}$,
\[
\frac{\partial}{\partial \tau}(\lSpaceDer{m}-\lambda^m \pSpaceDer{m})
    =-(\lSpaceDer{m}-\lambda^m \pSpaceDer{m}) \, ,
\]
 with
$\lSpaceDer{m}=\frac{\partial^m}{\partial s^m}\lSpace(s)$ calculated directly from
$\lSpace(s)$ on the loop by differentiation. 
In loop variables $\lSpace(s)$ we have,
\bea
	&& 
\frac{\partial^{m+1} \lSpace}{\partial s^m \partial \tau}-\lambda^m
\sum_{k=0}^m  
\frac{\partial f}{\partial \pSpaceDer{k}} \cdot \frac{\partial}{\partial \tau}
\frac{\partial^{k} \lSpace}
{\lambda^k \partial s^{k}}-m \lambda^{m-1} \lSpaceDer{m} \frac{\partial \lambda}
{\partial \tau}
	\continue
	&& \qquad\qquad\qquad\qquad\qquad
=\lambda^m \pSpaceDer{m}-\lSpaceDer{m}  
\, , \label{gen}
\eea
 where $\pSpace=\pSpaceDer{0}$ and $\lSpaceDer{k}=\frac{\partial^k 
\lSpace}{\lambda^k \partial^k s},\,k=1,\cdots,m-1$ are assumed.
Conventionally, 
\refeq{gf} is
converted to a system of $m d$ first order differential equations, 
whose discretized derivative (see \refeq{dd} below) are banded matrices
with band width of $5md$. Using \refeq{gen},
we only need $d$ equations for the same accuracy and the corresponding band
width is $(m+4)d$. The computing load has been greatly reduced, the 
more so the larger $m$ is. 
Nevertheless, choice of a good initial loop guess and
visualization of the dynamics are always aided by a plot of 
the orbit in the full $md$-dimensional phase space, where loops cannot 
self-intersect and topological features of the flow is exhibited 
more clearly.

\section{Implementation of \descent}
\label{sec:nm}

    As the loop points satisfy a periodic boundary condition, 
it is natural to employ truncated discrete Fast Fourier Transforms (FFT)
in numerical integrations of 
\refeq{bd3}.  Since we are
interested only in the final, stationary cycle $p$, the accuracy of the
fictitious time integration is not crucial; all we have to ensure is the smoothness of the loop 
throughout the integration.
The Euler integration with fairly large time steps $\delta \tau$ suffices.
The computationally most onerous step in implementation of
the {\descent} is the inversion of large matrix $\bar{A}$ in \refeq{bd3}.
When the dimension of the dynamical phase
space of \refeq{fl} is high, the inversion of
$\bar{A}$ needed to get $\frac{\partial \lSpace}{\partial \tau}$
takes most of the integration time, making the evolution extremely slow. 
This problem is partially
solved if the finite difference methods are used. The large matrix $\bar{A}$
then becomes sparse and the inversion can be done far more quickly.

\subsection{Numerical implementation}

    In a discretization of a loop, numerical
stability requires accurate  
discretization of  loop derivatives such as
 \[
\lVeloc_n \equiv
\left.\frac{\partial \lSpace}{\partial s}\right|_{\lSpace = \lSpace(s_n)}
	\approx (\hat{D}\lSpace)_n
\,.
\]
In our numerical work we use the four-point approximation\rf{brand03},
{\small 
\beq 
\hat{D} = 
\frac{1}{12h}
\!
\left( \begin{array}{ccccccccccc} 0&8&-1&&&\qquad
&&&&1&-8 \\ -8&0&8&-1&&\qquad &&&&&1 \\ 1&-8&0&8&-1&\qquad &&&&&\\
&&&&&\cdots&&&&&\\
&&&&&\qquad &1&-8&0&8&-1 \\
-1&&&&&\qquad &&1&-8&0&8 \\
8&-1&&&&\qquad &&&1&-8&0 
\end{array}\right)
\label{dd}
\eeq
}
where $h={2 \pi}/{N}$. 
Here, each entry represents a $[d\! \times\! d ]$ matrix, $8 \to 8\mathds{1}$, 
{\em etc.}, 
with blank spaces are filled with zeros.  
The two $[2d\! \times\! 2d]$ matrices
\[
M_1=\MatrixII{\mathds{1}}{-8\mathds{1}}{0}{\mathds{1}} \,, \qquad 
M_2=\MatrixII{-\mathds{1}}{0}{8\mathds{1}}{-\mathds{1}} \,,
\]
located at the top-right and bottom-left corners take care of 
the periodic boundary condition.

The discretized version
of \refeq{bd3} with a fictitious time Euler step $\delta \tau$ is
\beq
\MatrixII
 {\hat{\Mvar}}{\hat{v}}
 {\hat{a}}{0} 
  \VectorII {\delta \hat{x}}{\delta \lambda} 
=\delta \tau \VectorII{\lambda \hat{v}-\hat{\lVeloc}} {0}
\,, 
\label{mform} 
\eeq
where 
\bea
\hat{\Mvar} &=& \hat{D}+\mathrm{diag}[\Mvar_1,\Mvar_2,\cdots,\Mvar_N]
\,,
\nnu
\eea
with $\Mvar_n=\Mvar(\lSpace(s_n))$ defined in \refeq{doa}, and
\bea
\hat{v} &=& (v_1,v_2,\cdots,v_N)
		\,,\qquad \mbox{ with } v_n=v(\lSpace(s_n))\,,
	\continue
\hat{\lVeloc} &=& (\lVeloc_1,\lVeloc_2,\cdots,\lVeloc_N) 
\,,\qquad \mbox{ with }  
\lVeloc_n=\lVeloc(\lSpace(s_n)) \,,
\nnu
\eea
 are the two vector fields that we want to match everywhere along the loop.
$\hat{a}$ is an $Nd$ dimensional row vector which imposes the constraint on
the coordinate variations 
$\delta\hat{x}=(\delta \lSpace_1,\delta \lSpace_2,\cdots,\delta \lSpace_N)$.
 The discretized {\descent} \refeq{mform} is
an infinitesimal time step variant of the multipoint (Poincar\'{e} section) 
shooting equation for flows\rf{ChristiansenDasBuch}. 
In order to solve for the deformation of the loop coordinates and period, $\delta \hat{x}$ 
and $\delta \lambda$,
we need to invert the $[(N\, d+1)\! \times\! (N\, d+1)]$ matrix on the left hand side 
of \refeq{mform}. 

    In our numerical work, this matrix is inverted using the banded LU decomposition 
on the embedded
band-diagonal matrix, and the Woodbury formula\rf{nr} on the cyclic, border
terms.  The LU decomposition takes most of the computational
time and considerably slows down the fictitious time integration. We speed 
up the integration
by a new inversion scheme which relies on 
the smoothness of the flow in the loop space. 
It goes as follows. Once we have the
LU decomposition at one step, we use it to approximately invert the
matrix in the next step, with accurate inversion achieved
by the iterative approximate inversions\rf{nr}. In
our applications we find that a single LU decomposition can be used 
for many $\delta \tau$ evolution steps.
The further we go, the more
iterations at each step are needed to implement the inversion. After
the number of such iterations exceeds some  given fixed maximum
number, we perform another LU decomposition and proceed as before.  The
number of integration steps following one decomposition is an
indication of the smoothness of the evolution, and we adjust accordingly the
integration step size $\delta \tau$: the greater the number, the bigger the
step size. As the loop approaches a cycle, the
evolution becomes so smooth that the step size can be brought all the
way up to $\delta \tau = 1$, the full undamped Newton-Raphson iteration step. 
In practice, one can 
start with a small but reasonable number of points,
in order to get a coarse solution of relatively low accuracy. After
achieving that, the refined guess loop can be constructed
by interpolating more points, and proceed with 
for a more accurate calculation in which  $\delta \tau$ can be set as 
large as the full Newton step $\delta \tau = 1$, 
recovering the rapid quadratic convergence
of the Newton-Raphson method.

    It is essential that
the smoothness of the loop is maintained throughout
the calculation. We monitor the smoothness by checking the
Fourier spectrum of $\lSpace(\cdot,\tau)$. An unstable difference scheme
for loop derivatives might lead to unbounded sawtooth oscillations\rf{ns}. 
A heuristic local linear stability
analysis (described in \rf{lls}) indicates that our scheme is stable, and that 
the high frequency components do not generate instabilities.

\subsection{Initialization and convergence}
\label{subsec:init}

    As in any other method, a qualitative understanding of the dynamics is 
a prerequisite to successful cycle searches. We start by numerical 
integration with the dynamical system \refeq{fl}. Numerical experiments reveal
regions where a trajectory spends most of its life, giving us
the first hunch as to how to initialize a loop. We take the FFT of some nearly 
recurred orbit segment and keep only the lowest
frequency components. The inverse Fourier transform back to the phase
space yields a smooth loop that we use as our initial guess.
Since any generic orbit segment is not closed and might exhibit large 
gaps, the Gibbs phenomenon can take the initial loop so constructed
quite far away from the region of interest. We deal with this problem by 
manually deforming the orbit segment into a closed loop
before performing the FFT. Searching for longer
cycles with multiple circuits requires more delicate initial
conditions. The hope is that a few short cycles can help us establish 
an approximate symbolic dynamics, and guess for longer cycles 
can be constructed by cutting and glueing 
the short, known ones. For low dimensional systems, such methods yield 
quite good systematic initial guesses for longer cycles \rf{ks}.

An alternative way to initialize the search is by utilizing adiabatic
deformations of dynamics, or the homotopy
evolution\rf{cconley}. If the dynamical system \refeq{fl} depends on a 
parameter $\mu$, 
short cycles might survive as $\mu$ varies passing through a family of 
dynamical systems, giving in the process birth to
new cycles through sequences of bifurcations. Most short
unstable cycles vary little for small changes
of $\mu$. So, a cycle existing for parameter value $\mu_1$ 
can be chosen as the initial
trial loop for a nearby cycle surviving a small change $\mu_1 \rightarrow \mu_2$.
In practice, one or two iterations often suffice to find the new cycle.

A good choice of the initial loop significantly expedites the
computation, but there are more reasons why good initial loops are crucial.
 First of all, if we break the translational invariance by
imposing a constraint such as $\lSpace_1(s_1,\tau)=c$, we have to 
make sure that both the initial loop and
the desired cycle intersects this Poincar\'{e} plane. Hence,
the initial loop cannot be wildly different from the desired cycle. 
Second, in view of 
\refeq{cc}, the loop
always evolves towards  a local minimum of the {\costFct}al \refeq{funal}, 
with discretization points moving along the $\lVeloc-\lambda v$  
fixed direction, determined by the
initial condition. If the local minimum corresponds to a zero of
the {\costFct}al, we obtain a true cycle of the dynamical flow \refeq{fl}.
However, if the value of {\costFct}al is not equal to zero at the
minimum while the gradient is zero, \refeq{mform} yields a singular matrix
$\hat{\Mvar}$. In such cases the search has to be abandoned and restarted with a
new initial loop guess. In the periodic orbit searches of \refsect{sec:applt}
starting with blind
initial guesses (guesses unaided by a symbolic dynamics partition), such local 
minima were encountered in about $30\%$ of cases.

\subsection{Symmetry considerations}

    The system under consideration often possesses certain 
symmetries. If this is the case, the symmetry should be 
both be feared for possible marginal eigen-directions, and be embraced as a guide 
to possible simplifications of the numerical calculation. 

If the dynamical system equations \refeq{fl} are invariant under a discrete symmetry, 
the concept of fundamental domain\rf{CvitaEckardt,DasBuch} can be
utilized to reduce the length of the initial loop when searching for a 
cycle of a given symmetry.
In this case, we need discretize only an irreducible segment of the
loop, decreasing significantly the dimensionality of the loop representation. Other 
parts of the loop are replicated by symmetry operations,
with the full loop tiled by copies of the fundamental domain segment.
 The boundary conditions are not periodic any longer,
but all that we need to do is modify the cyclic terms. Instead of using
$M_1$ and $M_2$ in \refeq{dd}, we use $M_1 Q$ and $M_2 Q^{-1}$, where $Q$ is 
the relevant symmetry operation that maps 
the fundamental segment to the neighbor that precedes it. 
In this way, a fraction of the points represent the
cycle with the same accuracy, speeding up the search considerably.

    If a continuous symmetry is present, it may complicate the situation at
first glance but becomes something that we can take advantage of after careful
checking. 
For example, for a Hamiltonian system unstable cycles may form continuous
families\rf{henonrtb2,condisfam1}, with one or more members of a family belonging
to a given constant energy surface. In order to cope with the marginal eigendirection
associated with such continous family,
we search for a cycle on a particular energy surface by
replacing the last row of equation \refeq{mform} by an energy shell 
constraint\rf{ChristiansenDasBuch}. We
put one point of the loop, say $\lSpace_2$, on the constant energy
surface $H(\lSpace)=E$, and impose the constraint
$\bigtriangledown H(\lSpace_2) \cdot \delta \lSpace_2=0$, so as to 
keep $\lSpace_2$ on the surface for all $\tau$. The integration of 
\refeq{bd3} then automatically
brings all other loop points to the same energy surface. Alternatively, we
can look for a cycle of given fixed period $T$ by fixing $\lambda$ and
dropping the constraint in the bottom line of 
\refeq{mform}. These two approaches are conjugate to each other, both
needed in applications. 
 In most cases, they are equivalent. One exception is
the harmonic oscillator for which the oscillations have identical
period but different energy. Note that in both cases the translational
invariance is restored, as we have discarded the Poincar\'{e} section condition
of \refsect{sect:marg}. As explained in\rf{decar2}, this causes no trouble in
numerical calculations.

\section{Applications}
\label{sec:applt}

    We have checked that the iteration of \refeq{mform} yields quickly and robustly
the short unstable cycles for standard models of low-dimensional dissipative flows such as 
the R\"{o}ssler system\rf{ross}. A more daunting challenge are searches for cycles
in Hamiltonian flows, and searches for spatio-temporally periodic solutions of PDEs.
In all numerical examples that follow, the convergence condition is $\costF< 10^{-5}$.

\subsection{H\'enon-Heiles system and restricted three-body problem}

First, we test the 
Hamiltonian version of the \descent\ derived in \refSect{sec:ext}
by applying the method to two Hamiltonian systems, both with two degrees of freedom. 
In both cases, our initial loop guesses are rather arbitrary
combinations of trigonometric functions. Nevertheless, the 
observed convergence is fast. 

\begin{figure}[t] 
(a)~\includegraphics[width=1.8in]{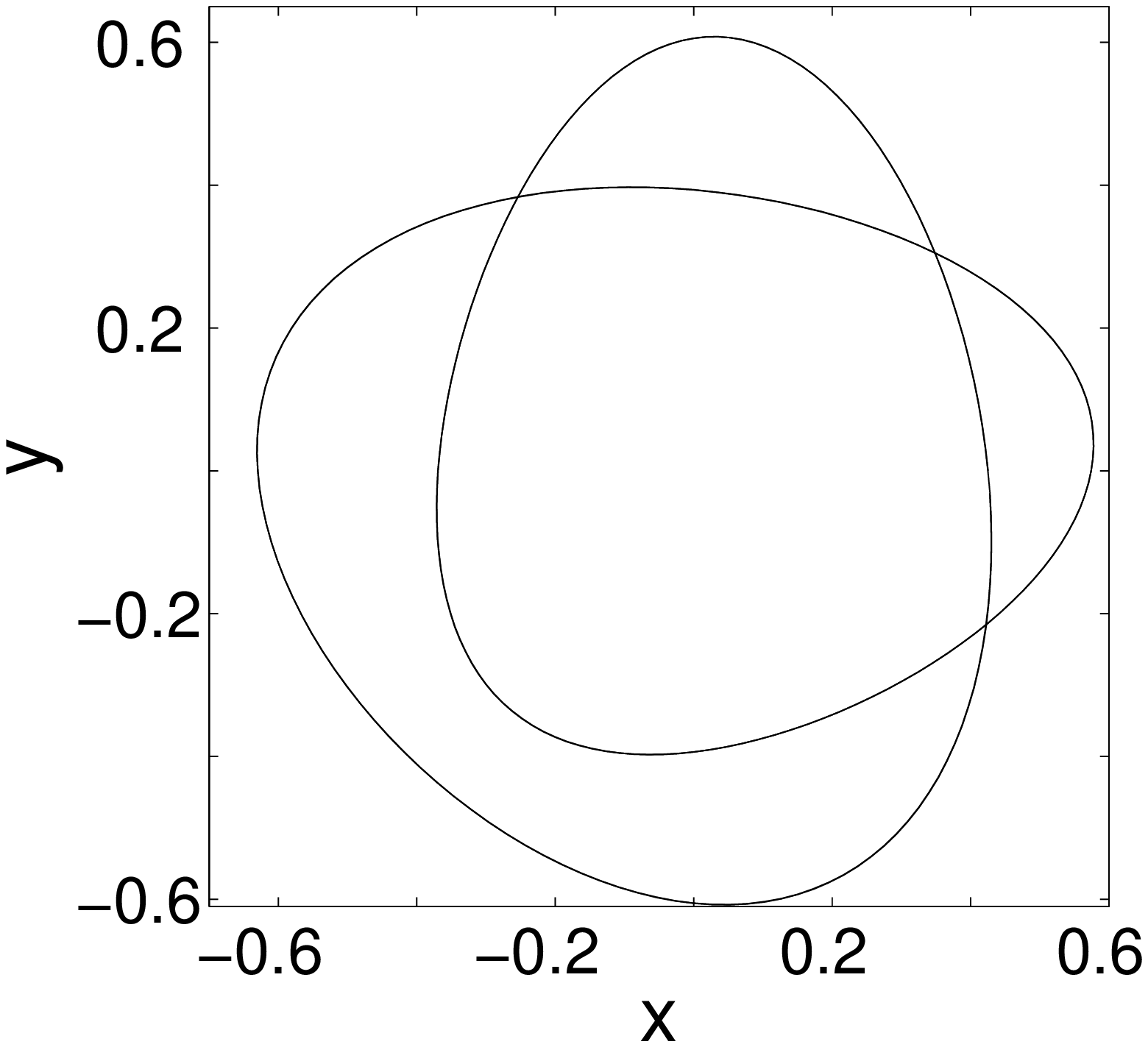}
(b)~\includegraphics[width=1.8in]{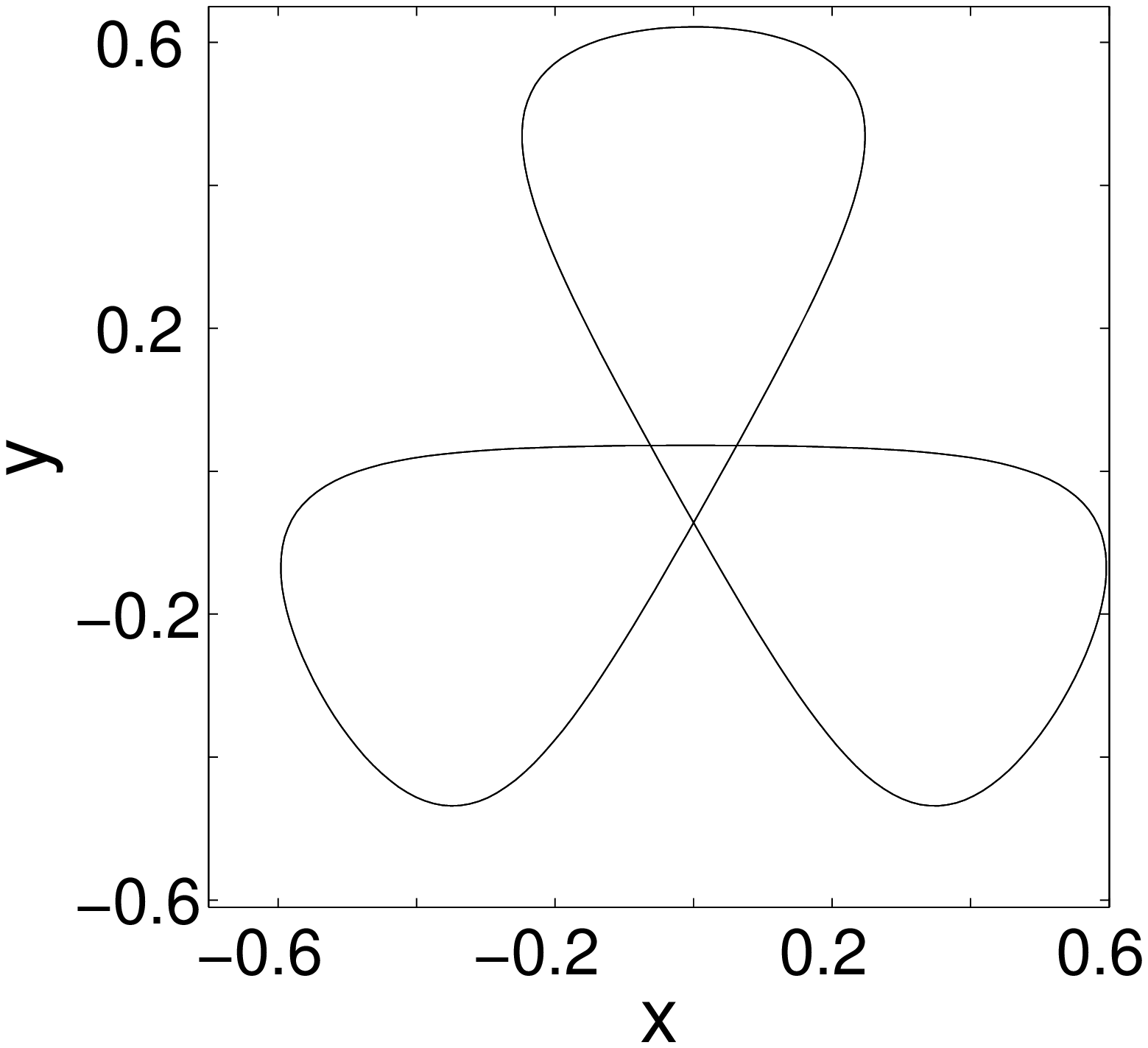}
(c)~\includegraphics[width=1.8in]{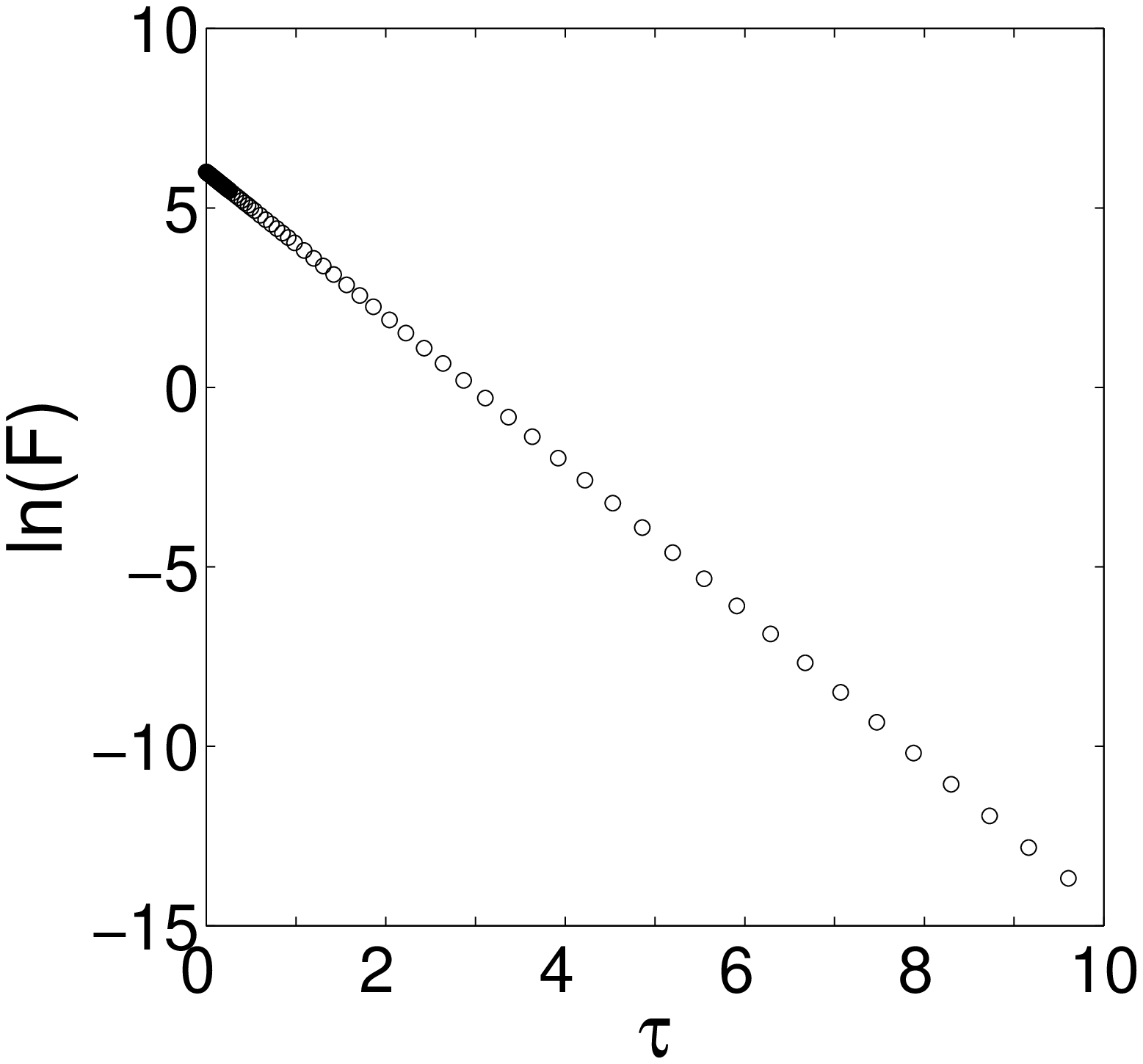}
\caption{
 The H\'enon-Heiles system in a chaotic region:
 (a) An initial loop $L(0)$, and 
 (b) the unstable periodic orbit $p$ of period $\period{}=13.1947$ reached by 
the {\descent} \refeq{hfl}.
 (c) The exponential decrease of the \costFct, $\ln (\costF) \approx -2.0502\,\tau+6.0214$.
    }
\label{f:hh}
\end{figure}

The H\'enon-Heiles system\rf{hhsys} is a standard model in celestial
mechanics, described by the Hamiltonian
 \begin{equation}
H=\frac{1}{2}(p_x^2+p_y^2+x^2+y^2)+x^2 y-\frac{y^3}{3} 
\,.
\end{equation}
 It has a time reversal symmetry and a three-fold discrete spatial symmetry.
 \refFig{f:hh} shows a typical application
of \refeq{hfl}, with the \descent\ search restricted to the configuration space. 
The initial loop, \reffig{f:hh}(a),  is a rather coarse initial guess. We fix arbitrarily 
the scaling $\lambda=2.1$, {\ie}, we search for a cycle $p$ of 
the fixed period $\period{p}=13.1947$, with no constraint on the energy.
\refFig{f:hh}(b) shows the cycle found by the \descent, with energy $E=0.1794$,
and the full discrete symmetry of the Hamiltonian. 
This cycle persists adiabatically for a small range of values of $\lambda$; 
with $\lambda$ changed
much, the \descent\ takes the same initial loop into other cycles. 
\refFig{f:hh}(c) verifies that the {\costFct}al
  $\costF$ decreases exponentially with slope -2 throughout the  $\tau=[0,10]$ 
integration interval, as predicted by \refeq{cc}.
The points get more and more sparse as $\tau$ increases, because
our numerical implementation adaptively chooses bigger and bigger step 
sizes $\delta \tau$.  

\begin{figure}[t] 
(a)~\includegraphics[width=2.0in]{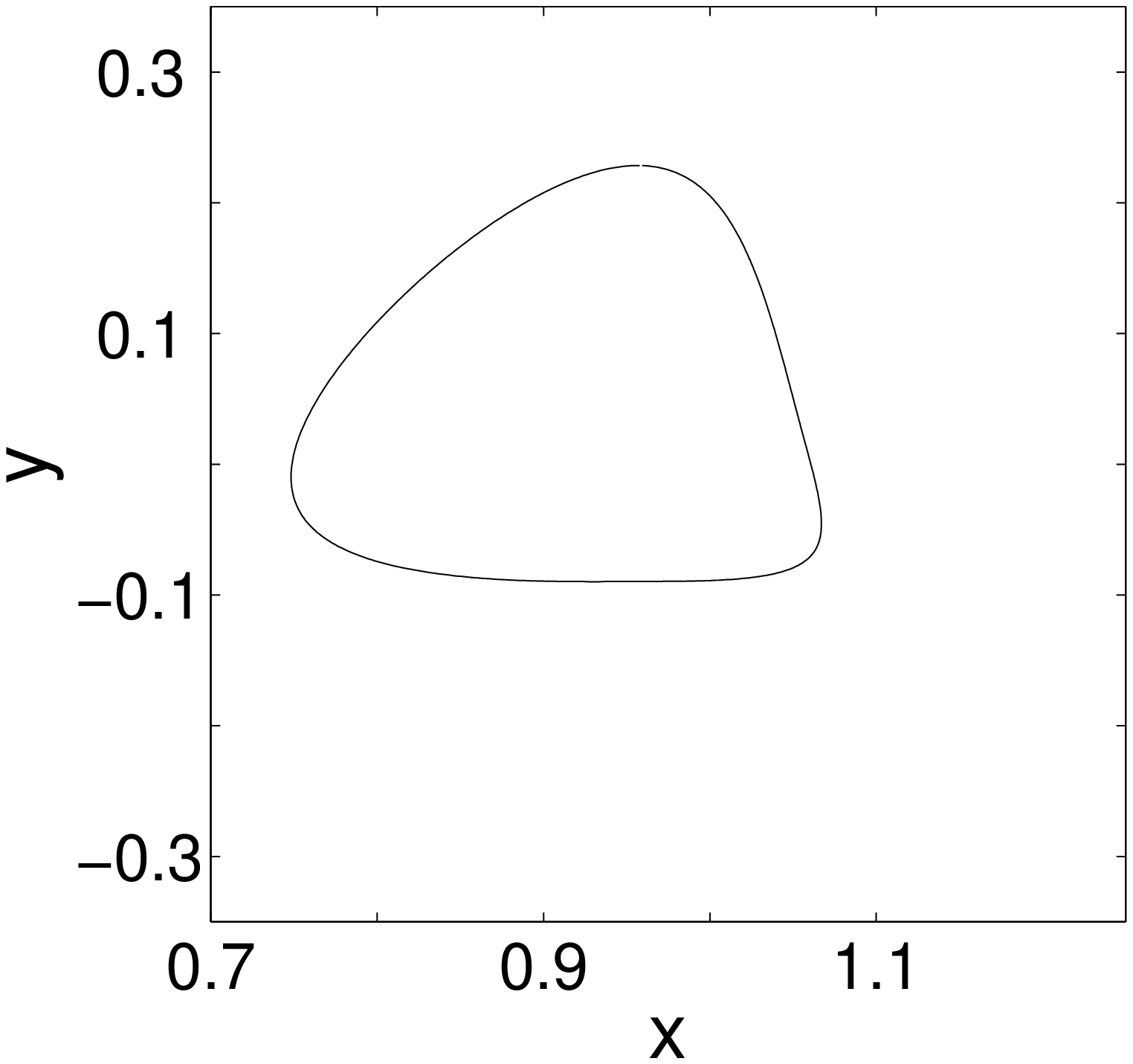}
\hspace{0.2in}
(b)~\includegraphics[width=2.0in]{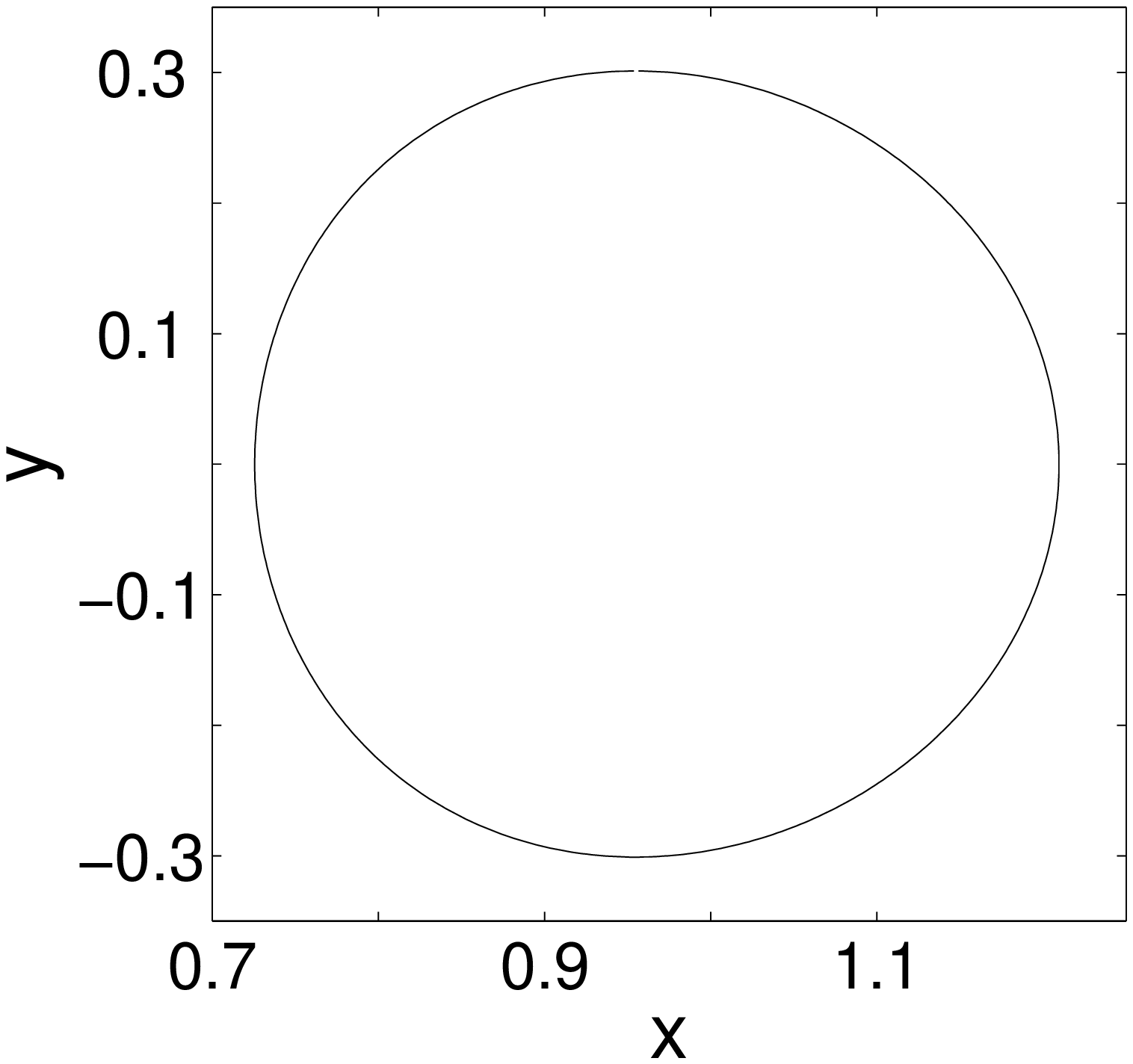}
\caption{
 (a) An initial loop $L(0)$, and 
 (b) the unstable periodic orbit $p$ of period $\period{p}=2.7365$ reached by 
the {\descent} \refeq{hfl}, the restricted
 three body problem \refeq{rtbeq} in the chaotic regime, $\mu=0.04$.
	 }
\label{rtb}
\end{figure}

 In the H\'{e}non-Heiles case, the accelerations $a_x,a_y$ depend only on the
configuration variables $x,y$. More generally, the accelerations could also
depend on $\dot{x},\dot{y}$. Consider as an example the equations of motion for the
restricted three-body problem\rf{rtb},
\begin{eqnarray}
\ddot{x} &=& 2\dot{y}+x
   -(1-\mu)\frac{x+\mu}{r_1^3}-\mu \frac{x-1+\mu}{r_2^3} \,,\nonumber \\
\ddot{y} &=& -2\dot{x}+y
   -(1-\mu)\frac{y}{r_1^3}-\mu \frac{y}{r_2^3} \label{rtbeq}
\,,
\end{eqnarray}
 where $r_1=\sqrt{(x+\mu)^2+y^2},\, r_2=\sqrt{(x-1+\mu)^2+y^2}$. These equations
describe the motion of a test particle in a rotating frame under the
influence of the gravitational force of two heavy bodies with masses $1$ and
$\mu \ll 1$ fixed at $(-\mu,0)$ and $(1-\mu,0)$ in the $(x,y)$ coordinate frame. The
stationary solutions of 
\refeq{rtbeq} are called the Lagrange points,
corresponding to a circular motion of the test particle in phase
with the rotation of the heavy bodies. The periodic solutions in the rotating 
frame correspond
to periodic or quasi-periodic motion of the test particle in the inertial
frame. \refFig{rtb} shows an initial loop and the cycle to
which it converges, in the rotating frame. Although the cycle looks simple,
the {\descent} requires advancing in small $\delta \tau$ steps in
order for the initial loop to converge to it. 
 
 In order to successfully
apply the Hamiltonian version of the {\descent} \refeq{hfl}, we 
have to ensure that the test
particle keeps a finite distance from the origin. If
a cycle passes very close to one of the heavy bodies, 
the acceleration can become so large that
our scheme of uniformly distributing the loop points in time might fail to
represent the loop faithfully. Another distribution scheme is required in
this case, for example, making the density of points proportional to the
magnitude of acceleration. 

\subsection{Periodic orbits of Kuramoto-Sivashinsky system}

   The Kuramoto-Sivashinsky equation arises as an amplitude equation
for interfacial instability in a variety of contexts\rf{ku,siv}. 
In 1-dimensional space, it reads
\begin{equation}
u_t=(u^2)_x-u_{xx}-\nu u_{xxxx}, \label{kseq}
\end{equation}
 where $\nu$ is a ``super-viscosity'' parameter which controls the rate of
dissipation and $(u^2)_x$ is the nonlinear convection term. 
As $\nu$ decreases, the system undergoes a series of bifurcations,
leading to increasingly turbulent, spatio-temporally chaotic dynamics.

 If we impose the periodic boundary condition
$u(t,x+2\pi)=u(t,x)$ and choose to study only the odd solutions
$u(-x,t)=-u(x,t)$, $u(x,t)$ can be expanded in a discrete spatial
Fourier series\rf{ks},
\begin{equation}
u(x,t)=i\sum_{k=-\infty}^{\infty} a_k(t) e^{ikx}, \label{expan}
\end{equation} 
where $a_{-k}=-a_k \in \mathbb{R}\,$. In terms of the Fourier components, 
PDE \refeq{kseq} becomes an infinite ladder of ODEs,
\begin{equation}
\dot{a_k}=(k^2-\nu k^4)a_k-k\sum_{m=-\infty}^{\infty}a_m a_{k-m} \,. \label{ksf}
\end{equation}
 In numerical simulations we work with the Galerkin truncations of the
Fourier series since in the neighborhood of the strange attractor the 
magnitude of $a_k$ decreases very fast with
$k$,  high frequency modes playing a negligible role
in the asymptotic dynamics. In this way Galerkin truncations reduce the
dynamics to a finite but large number of ODEs. We work with $d=32$ dimensions 
in our numerical 
calculations. In \refref{ks}, multipoint shooting
has been successfully applied to obtain periodic orbits close to the
onset of spatiotemporal chaos ($\nu=0.03$). In this regime, our method
is so stable that big time steps $\delta \tau$ can
be employed even at the initial guesses, leading to extremely fast convergence. 
We attribute this robustness to the simplicity of 
the structure of the attractor at high viscosity values.
%
\begin{figure}[t!]
(a)~\includegraphics[width=1.8in]{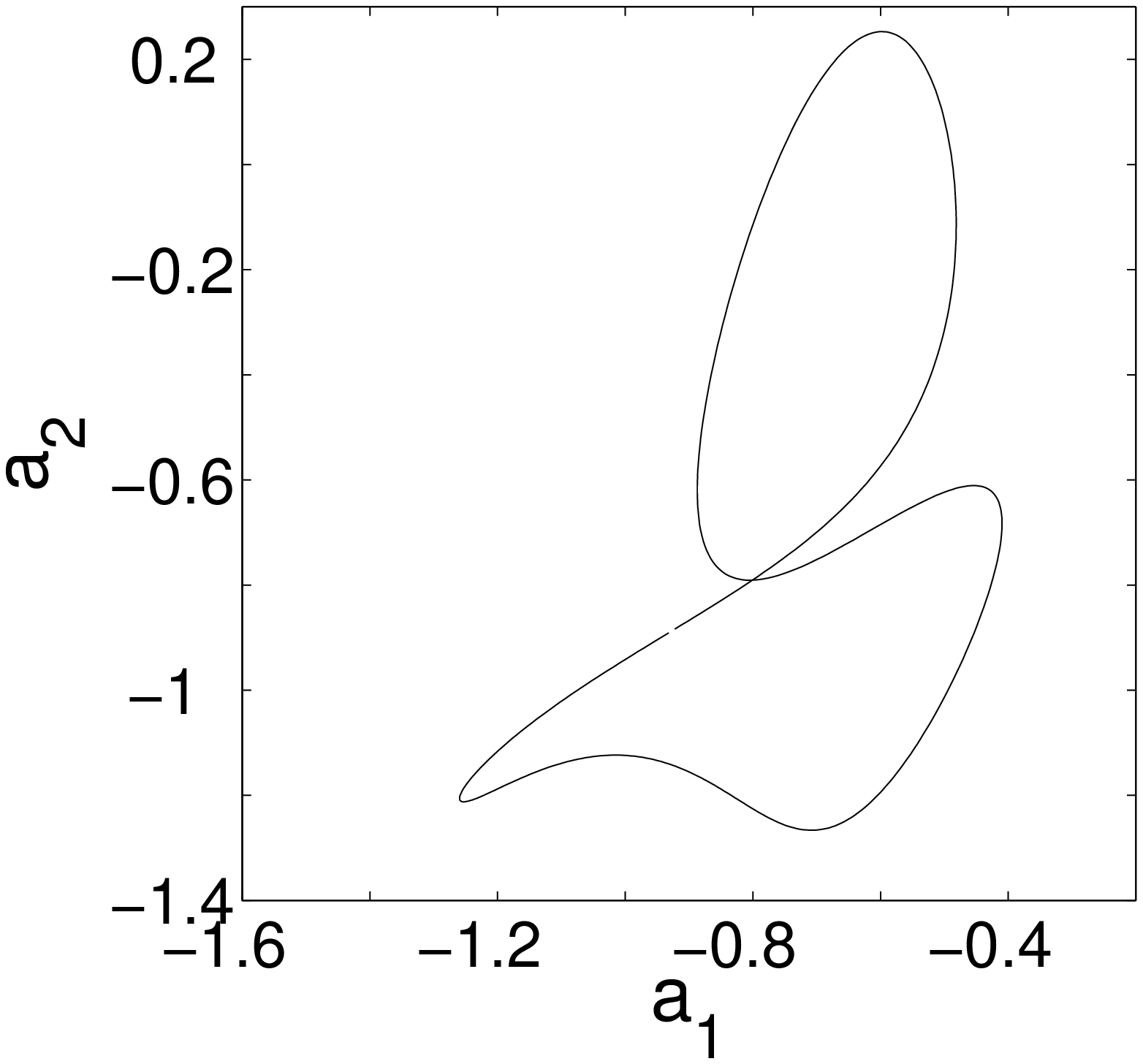}
\hspace{0.2in}
(b)~\includegraphics[width=1.8in]{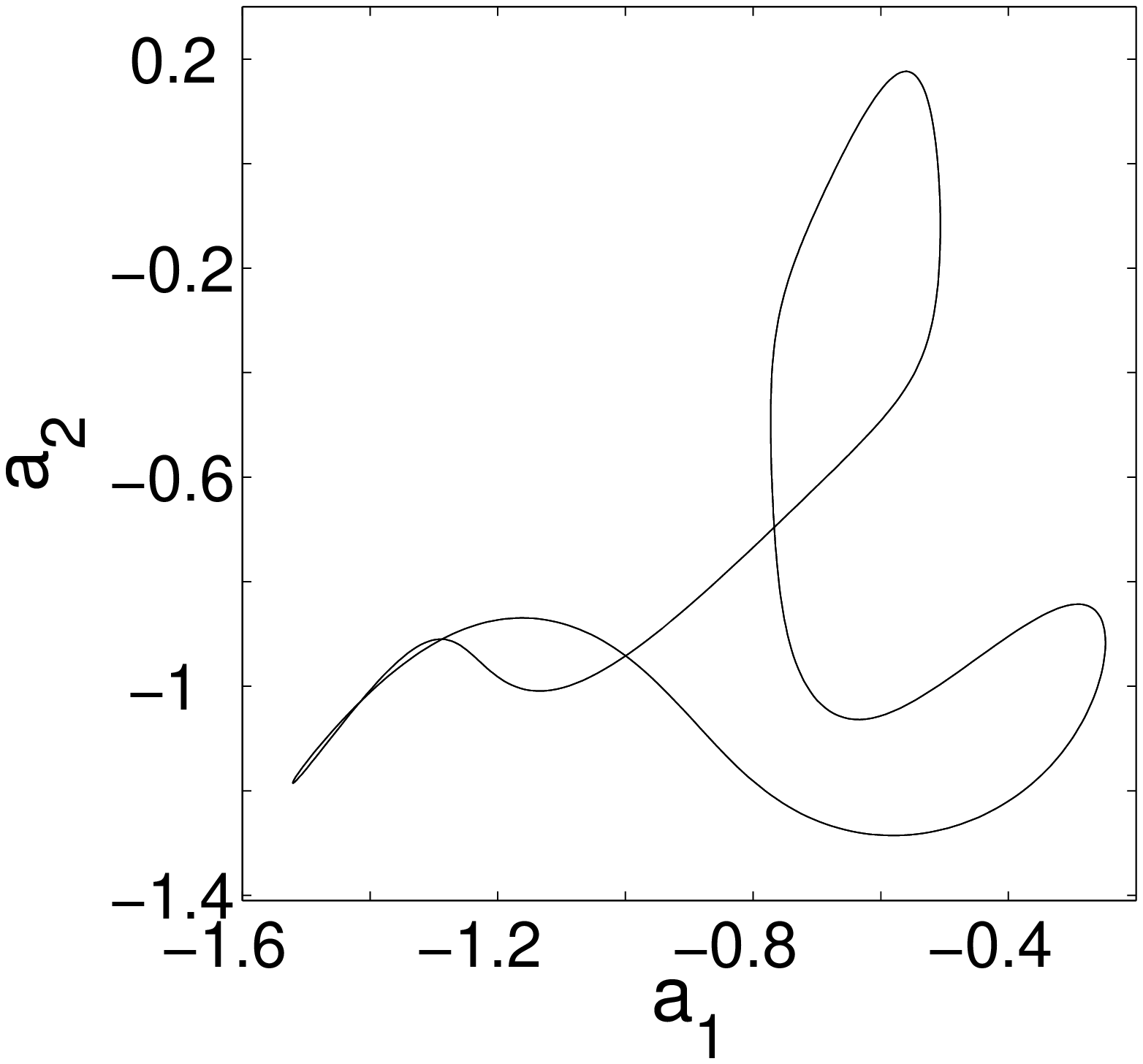}\\
(c)~\includegraphics[width=1.8in]{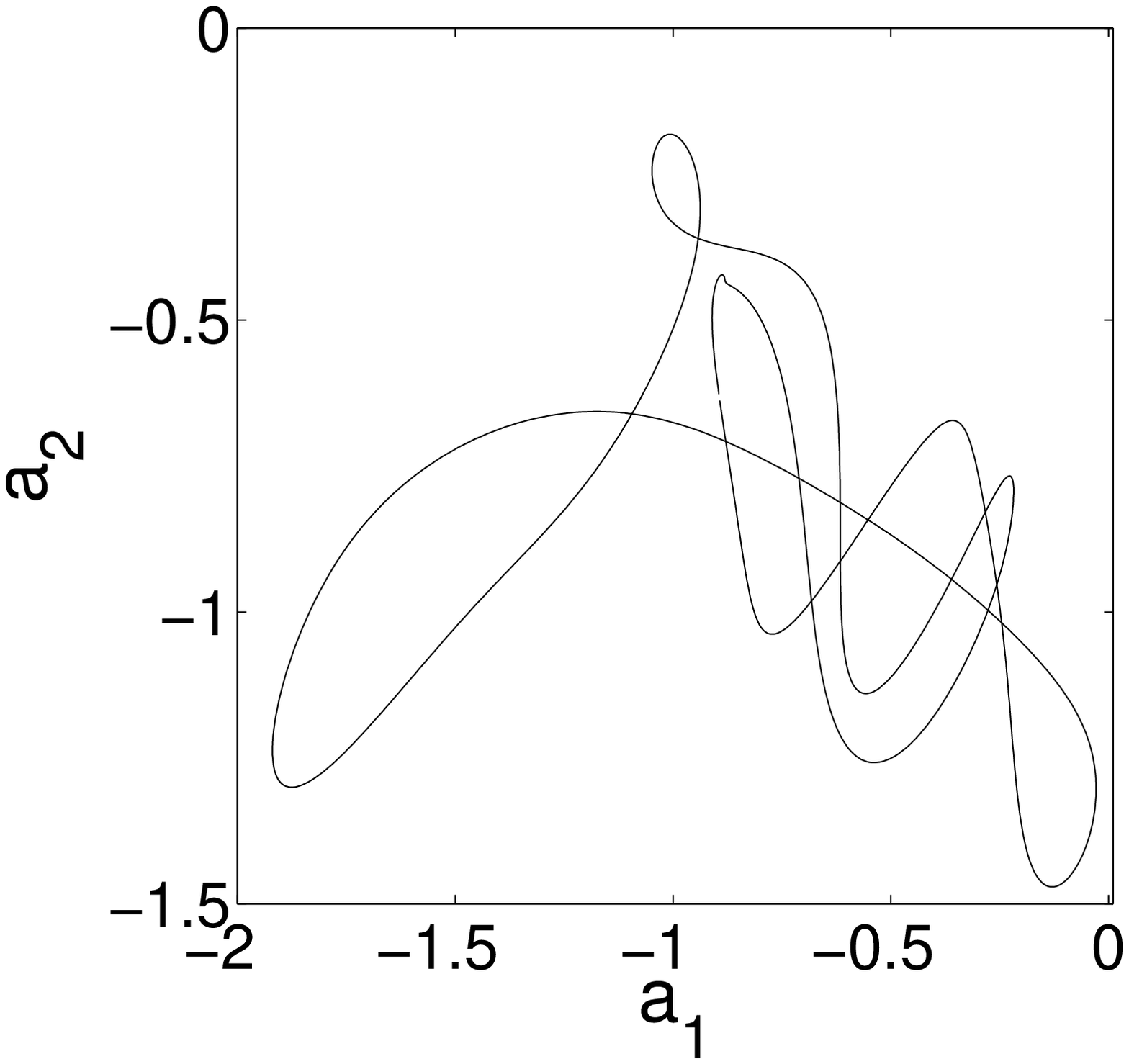}
\hspace{0.2in}
(d)~\includegraphics[width=1.8in]{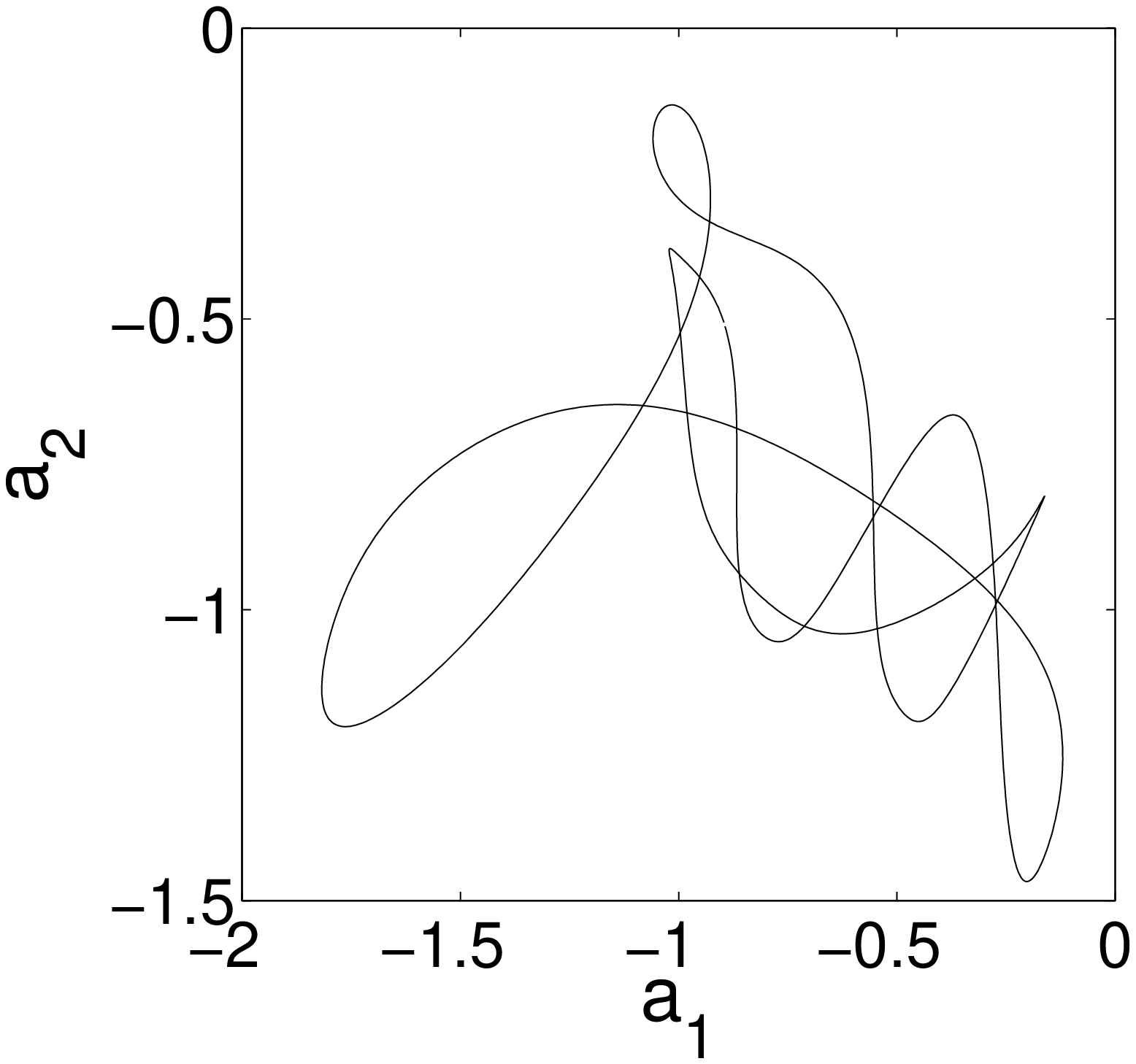}

\caption{
 The Kuramoto-Sivashinsky system in a spatio-temporally
 turbulent regime (viscosity parameter
 $\nu=0.015$, $d=32$ Fourier modes truncation). 
 (a) An initial guess $L_1$, and 
 (b) the periodic orbit $p_1$ of period $\period{1}=0.744892$ reached by the \descent.
 (c) Another initial guess $L_2$, and
 (d) the resulting periodic orbit $p_2$ of period $\period{2}=1.184668$.}
\label{f:ks1}
\end{figure}  

The challenge comes with decreasing $\nu$, with the dynamics turning more and
more turbulent.
Already at $\nu=0.015$, the system is moderately turbulent and 
the phase space portraits of the dynamics reveal a complex labyrinth of ``eddies''
of different scales and orientations. While the highly unstable nature of orbits
and intricate structure of the invariant set hinder
applications of conventional cycle-search routines, in this setting  
our variational method shines through. We design rather arbitrary initial loops
from numerical trajectory segments, and the calculation proceeds 
as before, except that now 
a small $\delta \tau$ has to be used initially to ensure numerical stability. 
Topologically different loops are very likely to result in different cycles,
while some initial loop guesses my lead to local nonzero minima of the {\costFct}al $\costF$.
As explained in   \refSect{sec:nm},  in such cases the method diverges, and
the search is restarted with a new inital loop guess.

\begin{figure}[t!] 
\centering
(a)~\includegraphics[width=2.2in]{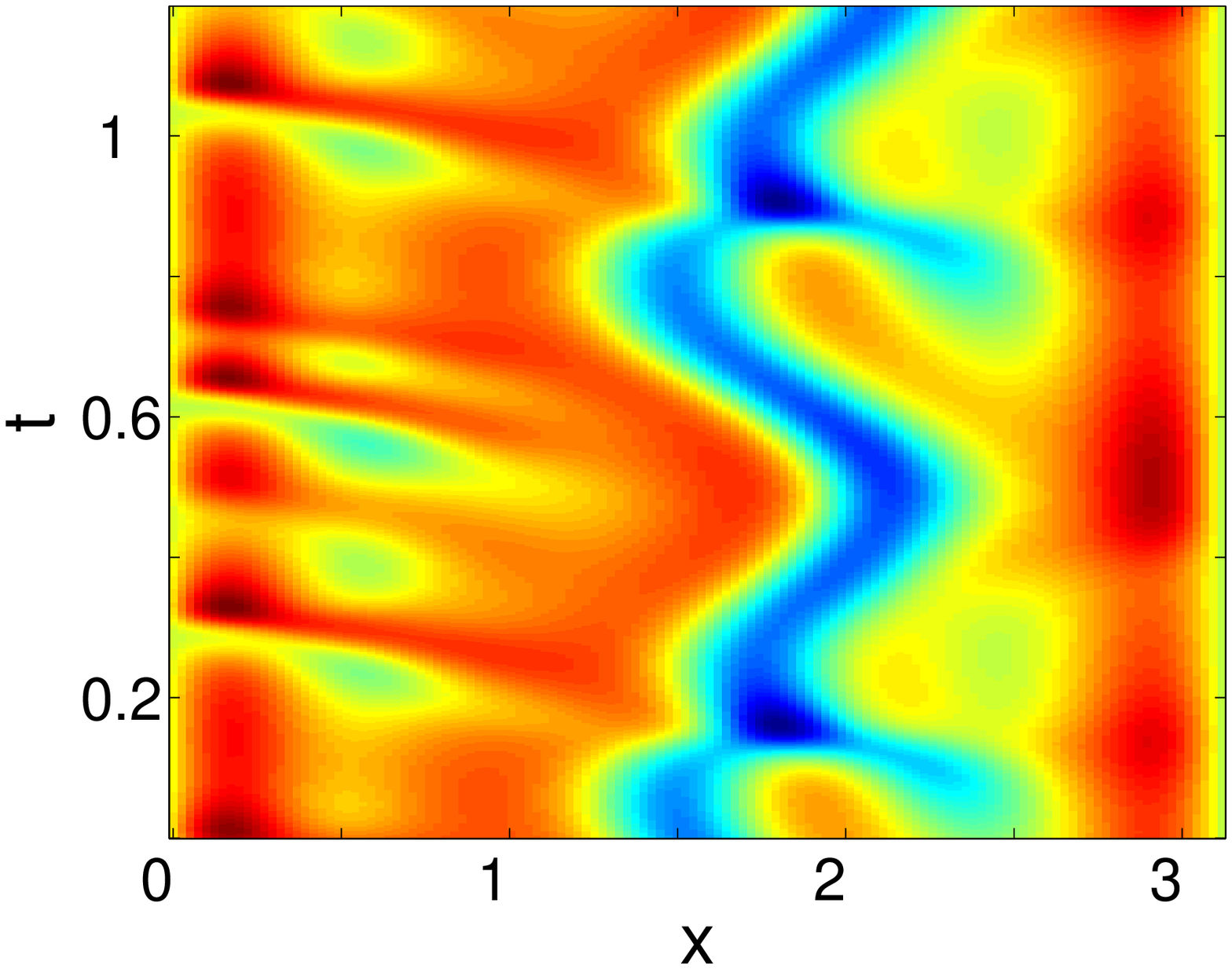}
\hspace{0.2in}
(b)~\includegraphics[width=2.2in]{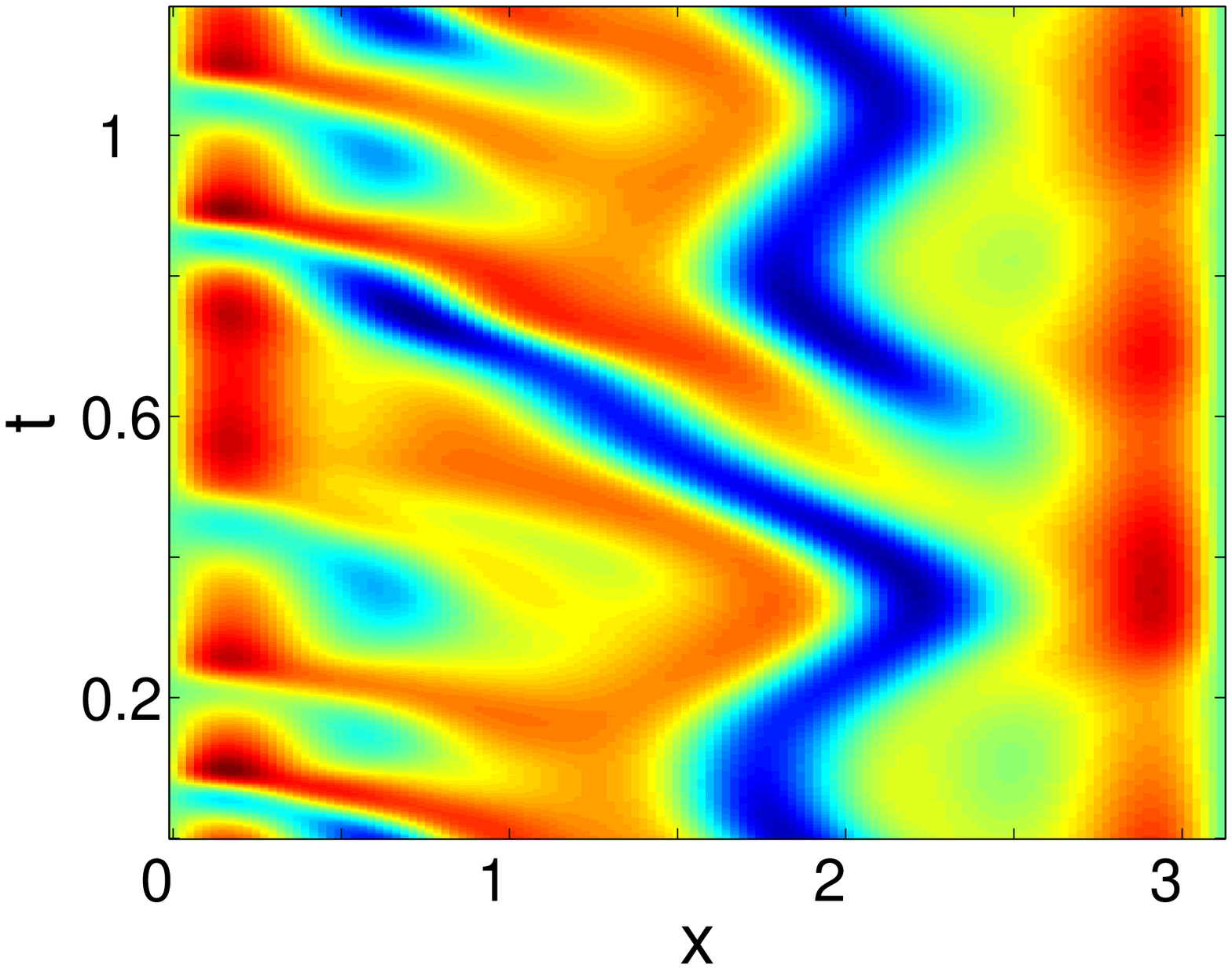}
\caption{
 Level plot of the space-time evolution $u(x,t)$ for the 
 two spatiotemporally periodic solutions of \reffig{f:ks1}: 
 (a) the evolution of $p_1$, with start of a repeat after the cycle period
$\period{1}=0.744892$, and  
(b) one full period $\period{2}=1.184668$ in
  the evolution of $p_2$.
	 }
\label{f:tevol}
\end{figure}

	Two initial loop guesses are displayed in
\reffig{f:ks1}, along with the two periodic orbits
detected by
the \descent. In discretization of the initial loops, each point has to be specified in
 all $d$~dimensions;
 here the coordinates $\{a_1,a_2\}$ are picked so that topological 
 similarity between initial and final loops is visually easy to identify.
 Other projections from $d=32$ dimensions to subsets of 2 coordinates 
 appear to make the identification harder, if not impossible.
In both calculations, we molded segments of typical trajectories into
smooth closed loops by the Fourier filtering method of
\refSect{sec:nm}. As the desired orbit
becomes longer and more complex, more sampling points are needed to 
represent the loop. We use $N=512$ points to represent the loop in the (a)-(b) case and
$N=1024$ points in the (c)-(d) case. The space-time evolution of $u(x,t)$ 
for these two unstable spatio-temporally periodic solutions is displayed 
in \reffig{f:tevol}.
As $u(x,t)$ is antisymmetric on $[-\pi,\pi]$, it suffices to 
display the solutions on the $x \in [0,\pi]$ interval.

\section{Discussion}
\label{sec:sum}

In order to cope with the difficulty of finding periodic orbits in
high-dimensional chaotic flows, we have devised the {\em \descent\ method},
 an infinitesimal step variant of
the damped Newton-Raphson method. 
Our main result is the PDE \refeq{bd3} which solves the variational problem of
minimizing the {\costFct}al
\refeq{funal}.
This equation describes the fictitious
time $\tau$ flow in the space of loops which decreases the {\costFct}al
at uniform exponential rate (see \refeq{cc}).
Variants of the method are presented for special classes of systems,
such as Hamiltonian systems. An efficient integration scheme for the PDE
is devised and tested on the Kuramoto-Sivashinsky system,
the H\'{e}non-Heiles system and the restricted three-body
problem.
     
Our method uses information from a large number of points in phase
space, with the global topology of the desired cycle protected by insistence on
smoothness and a uniform discretization of the loop. 
The method is quite robust in practice.

The numerical results presented here are only a proof of principle. We do not 
know to what periodic orbit the flow \refeq{bd3} will evolve for a 
given dynamical system and a given initial loop. 
Empirically, the flow goes toward the ``nearest'' periodic orbit,
with the largest topological resemblance. Each particular application will still require 
much work in order 
to elucidate and enumerate relevant topological structures. The hope is that the short 
spatio-temporally periodic solutions revealed by the {\descent} searches will serve as
the basic building blocks for systematic investigations of chaotic and perhaps even
``turbulent'' dynamics.

\begin{acknowledgments}

We would like to thank Cristel Chandre for careful reading of the
manuscript and numerous suggestions. 

\end{acknowledgments}         

\bibliography{../bibtex/nonlind}

\end{document}

%% file: loopDefs.tex
\newcommand{\fix}[1]			
             {{\color{magenta} #1 }}

\newcommand{\Loop}{L}
\newcommand{\costF}{F^2}  
\newcommand{\lSpace}{\tilde{x}}			
\newcommand{\lVeloc}{\tilde{v}}
\newcommand{\pSpaceDer}[1]{x^{(#1)}}
\newcommand{\lSpaceDer}[1]{\tilde{x}^{(#1)}}

\newcommand{\descent}{Newton descent}

\newcommand{\costFct}{cost function}	

\newcommand{\rf}      [1] {~\cite{#1}}
\newcommand{\refref}  [1] {ref.~\cite{#1}}

\newcommand{\refeq}   [1] {(\ref{#1})}

\newcommand{\reffig}  [1] {Fig.~\ref{#1}}

\newcommand{\refFig}  [1] {Figure~\ref{#1}}

\newcommand{\refsect} [1] {sect.~\ref{#1}}

\newcommand{\refSect} [1] {Sect.~\ref{#1}}

\newcommand{\beq}{\begin{equation}}
\newcommand{\eeq}{\end{equation}}
\newcommand{\ee}[1] {\label{#1} \end{equation}}
\newcommand{\bea}{\begin{eqnarray}}

\newcommand{\continue}{\nonumber \\ }
\newcommand{\nnu}{\nonumber}
\newcommand{\eea}{\end{eqnarray}}

\newcommand{\MatrixII}[4]{
   \begin{pmatrix} 
             {#1}  &  {#2} \cr
             {#3}  &  {#4} \cr    
   \end{pmatrix}     
                         }

\newcommand{\VectorII}[2]{
   \begin{pmatrix} 
             {#1} \cr
             {#2} \cr    
   \end{pmatrix}     
                         }

\newcommand{\ie}{{that is}}		

\newcommand{\jacobianM}{Jacobian matrix}	

\newcommand{\reals}{\mathbb{R}}

\newcommand{\pS}{{\cal M}}			
\newcommand{\Mvar}{{A}}	   	   
\newcommand{\jMps}{{J}}	   

\newcommand\flow[2]{{f^{#1}(#2)}}

\newcommand\pSpace{x}		

\newcommand\period[1]{{T_{#1}}}			
